\documentclass[aps,prl,twocolumn,superscriptaddress]{revtex4-1} 
\usepackage{graphicx}
\usepackage{amssymb, amsmath, amsthm, amsfonts, bm, mathrsfs, wasysym, txfonts, bbm, enumerate, color}

\begin{document}

\title{Tying knots in light fields  }
\author{Hridesh Kedia}
\email{hridesh@uchicago.edu}
\affiliation{University of Chicago, Physics department and the James Franck institute, 929 E 57$^{\rm th}$ st. Chicago, IL, 60605}
\author{Iwo Bialynicki-Birula}
\affiliation{Center for Theoretical Physics, Polish Academy of Sciences Al. Lotniko«w 32/46, 02-668 Warsaw, Poland}
\author{Daniel Peralta-Salas}
\affiliation{Instituto de Ciencias Matem\'aticas, Consejo Superior de Investigaciones Cient\'ificas, 28049 Madrid, Spain}
\author{William T.M. Irvine}
\affiliation{University of Chicago, Physics department and the James Franck institute, 929 E 57$^{\rm th}$ st. Chicago, IL, 60605}

\begin{abstract}
We construct a new family of null solutions to Maxwell's equations in free space whose field lines encode all torus knots and links. The  evolution of these null fields, analogous to a  compressible flow along the  Poynting vector that is both geodesic and shear-free,  preserves the topology of the knots and links. Our approach combines the Bateman and spinor formalisms for the construction of null fields with complex polynomials on $\mathbb{S}^3$. We examine and illustrate the geometry and evolution of the solutions, making manifest the structure of nested knotted tori filled by the field lines.

\end{abstract}
\maketitle

Knots and the application of mathematical knot theory to space-filling fields are deepening our understanding of a broad set of physical phenomena with examples in fluid dynamics, liquid crystals, optics, and topological field theories~\cite{ Chandrasekhar1956, Woltjer1958, Moffatt1969, Moffatt1981, Witten1989, Baez, Ricca1996, Faddeev1997, Manton2004a, Barenghi2007, Irvine2008, Babaev2009, Irvine2010,  Smalyukh2010, Dennis2010, Tkalec2011, Daniel2012}.
Knotted structures in hydrodynamical fields,  for example in  the current-guiding magnetic field lines of a plasma, or vortex lines of classical and quantum fluids, arise naturally as excitations that carry helicity, a conserved quantity that is a measure of the knottedness of the field.

Considerably more elaborate structures than their counterparts tied in shoelaces, to tie a knot in a field one has to worry not only about a single line, but about the entire field.
Moreover, dynamical and physical constraints inherent in each physical system can significantly constrain the freedom to construct arbitrary configurations and affect their stability.
Finding analytical expressions for dynamical knotted structures, is further complicated by the inherent nonlinearity of many  hydrodynamical systems.

\begin{figure}[!htb]
\includegraphics[width = \columnwidth]{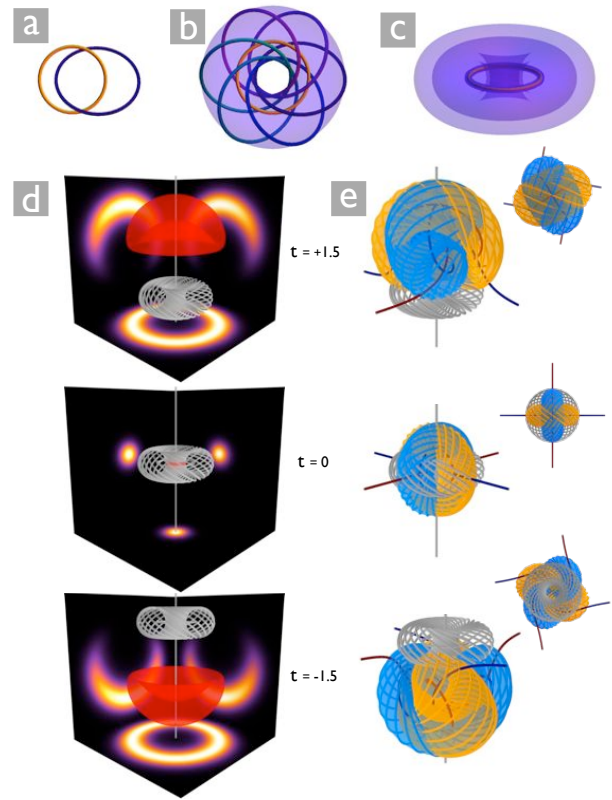}
\caption{{\bf Hopfion solution: field line structure (a-c) and time evolution (d-e)}.  {\bf a}: Hopf link formed by the circle at the core (orange) of the nested tori, and one of the field lines (blue). {\bf b}: A torus (purple) enclosing the core (orange), filled by mutually linked field lines. {\bf c}: Nested tori (purple) enclosing the core, on which the field lines lie. {\bf d}: Time evolution of the Poynting field lines (gray), an energy isosurface (red), and the energy density (shown via projections). {\bf e}: Time evolution of the electric (yellow), magnetic (blue) and Poynting field lines (gray).}
\label{Hopfion_fig}
\end{figure}%

Linked and knotted beams of light provide an opportunity to study a nontrivial topological structure in the setting of a linear field theory and furthermore provide a means of potentially transferring knottedness to matter.

A particularly elegant solution to Maxwell's equations re-discovered and studied in several contexts \cite{Bateman1915,Trautman1977,Ranada1989,Ranada1995,Urbantke2003,Bialynicki-Birula2004,Irvine2008}, is shown in Fig.~\ref{Hopfion_fig}. This solution encodes the simplest topologically non-trivial configuration: Hopf links, interlocked to form a Hopf fibration that fills space with linked circles. Remarkably, its topological structure is preserved in time. It can be constructed in many ways, using complex scalar maps, spinors, twistors and shall be referred to as the Hopfion solution for the rest of the letter.

A natural question is: do Maxwell's equations allow for other propagating solutions encoding more complex knotted configurations than a pair of linked rings?  Attempts at generalizing Hopfions to torus knots \cite{Irvine2008,Irvine2010,Trueba2010} showed that it is possible to construct such solutions at an instant in time, but their structure is not preserved \cite{Irvine2010}, and unravels in time, leaving open the problem of designing generalized solutions that encode non-trivial knots and whose entire topological structure is preserved in time. Beyond Maxwell's equations, the more general problem of finding explicit solutions to dynamical flows 
 which embody  persistent knots 
 has also remained open. 

From a hydrodynamic perspective \cite{Irvine2010,HDW}, if the electric and magnetic fields are everywhere perpendicular and of equal magnitude (the field is null), then the evolution of the field can be understood as a smooth flow that carries the field lines with it, guaranteeing their topological preservation.
In this letter we show how constructions based on Bateman's method \cite{Bateman1915} or equivalently a spinor formalism can lead to novel null solutions that encode knots of different kinds, and evolve in a smooth manner like the Hopfion solution.  The combination of a null electromagnetic field formalism with a topological construction, leading to a family of knotted null solutions is the central result of this paper.

{\it The Hopfion solution --}
The Hopfion electromagnetic field (Fig.~\ref{Hopfion_fig}) is a light-beam-like propagating solution, with the curious property that all electric and magnetic field lines are closed loops and any two electric (or magnetic) field lines are linked \cite{Ranada1989,Ranada1990,Ranada1995,Irvine2008}. At time t=0 the electric, magnetic and Poynting field lines have identical structure (that of a Hopf fibration), oriented in space so that they are mutually orthogonal to each other.

As the solution evolves, the electric and magnetic field line configurations deform smoothly, as if transported by a  fluid flow, and the Poynting field line structure moves, undeformed, in the z direction. This smooth topology-preserving time evolution of the field lines is characteristic of null electromagnetic fields \cite{Irvine2010,HDW} and is best interpreted in terms of the Poynting field structure.

By Robinson's theorem \cite{Robinson1961}, there is a shear-free family of light rays (a null GSF (geodetic, shear-free) congruence) underlying each null electromagnetic field that, in flat space-time, is given explicitly by the normalized Poynting field. The time evolution can then be thought of as fluid-like, with each parcel of fluid transported along the world lines of these rays. While the packets of electromagnetic field follow these straight line trajectories (geodesics in free space) the angles between $\mathbf{E}$ and $\mathbf{B}$ are preserved (and thus the fluid evolution is shear-free). In case of the Hopfion solution (for which the associated null GSF congruence is the Robinson congruence), the instantaneous structure of the Poynting field is sufficient to determine these straight line trajectories, even though it evolves with time (by a rigid translation in the z direction, as shown in Fig.~\ref{Hopfion_fig}).

The null condition makes the design of a knotted magnetic or electric field a problem of engineering {\it three} fields, imposing a strong geometric constraint on the possible topological configurations of field lines. As the construction of a triplet of mutually orthogonal fields, that remains orthogonal under time evolution is central to the problem of constructing knotted beams of light, we start with the formalisms developed for the construction of null fields and seek to construct knotted structures within them.

Null electromagnetic fields have a rich history, from the early construction by Bateman   \cite{Bateman1915} to Robinson's theorem \cite{Robinson1961} and Penrose's twistor theory \cite{Penrose1967}.
We now briefly summarize two methods for their construction.

{\it Bateman's construction -- }
Bateman \cite{Bateman1915} constructs all null electromagnetic fields associated with a given underlying null GSF congruence, starting from two complex scalar functions of space-time. 
Hogan \cite{Hogan1984}, has shown that all null electromagnetic fields can be constructed using Bateman's method.

According to Bateman's construction, given a pair of complex scalar functions of space-time $(\alpha, \beta)$ which satisfy:
\begin{equation}
{\mathbf\nabla}\alpha\times{\mathbf\nabla}\beta=\mathrm{i}\left(\partial_{t}\alpha{\mathbf\nabla}\beta
-\partial_{t}\beta\,{\mathbf\nabla}\alpha\right)\label{alphabeta_def}
\end{equation}
there is a corresponding electromagnetic field:
\begin{equation}
\mathbf{F}=\nabla\alpha\times\nabla\beta \label{emdef_bateman_basic}
\end{equation}
where $\mathbf{F}=\mathbf{E} + \mathrm{i}\mathbf{B}$, is known as the Riemann-Silberstein vector \cite{Bialynicki-Birula1996}. This field is null (both invariants vanish),
\begin{equation}
\mathbf{E}\cdot\mathbf{B}=0\,,\quad \mathbf{E}\cdot\mathbf{E}-\mathbf{B}\cdot\mathbf{B}=0 \label{null_cond}
\end{equation}
since the scalar product $\mathbf{F}\cdot\mathbf{F}$ is zero. For the null solutions generated by (\ref{emdef_bateman_basic}) to be non-trivial, the following conditions must be satisfied: 
\begin{align}\label{eq:nontrivial}
\partial_t\alpha\left((\partial_t\alpha)^2-({\mathbf\nabla}\alpha)^2\right)=0\,, \; \partial_t\beta\left((\partial_t\beta)^2-({\mathbf\nabla}\beta)^2\right)=0
\end{align}

Each pair $(\alpha,\beta)$ satisfying (\ref{alphabeta_def}) generates a whole family of fields because any vector field of the form:
\begin{equation}
\mathbf{F}=h(\alpha,\beta)\,{\mathbf\nabla}\alpha\times{\mathbf\nabla}\beta
={\mathbf\nabla}f(\alpha,\beta)\times{\mathbf\nabla}g(\alpha,\beta), \label{hfg_bateman}
\end{equation}
where $h:=\partial_\alpha f\,\partial_\beta g-\partial_\beta f\,\partial_\alpha g$ and $f,g$ are arbitrary holomorphic functions of $(\alpha,\beta)$, also gives rise to a null electromagnetic field. Note that all fields constructed in this way have, by construction, the same underlying null GSF congruence $(|\mathbf{E}\times\mathbf{B}|,\mathbf{E}\times\mathbf{B})$. 

This is made manifest when the above null field and its associated null GSF congruence are expressed in the equivalent language of spinors. In this formalism \cite{penrose1987spinors}, a null congruence $\xi^{\mu}$ is constructed from a spinor field $\xi_{A}$, and a null electromagnetic field $F_{\mu\nu}$ is constructed from a  symmetric spinor $\Phi_{AB}$ as:
\begin{align*}
\xi^{\mu} = g^{\mu AA'}\;\xi_{A}\;\bar{\xi}_{A'}\, ; \, 
F^{\mu\nu} = g^{\mu AA'}g^{\nu BB'}\left(\Phi_{AB}\,\epsilonup_{A'B'}
+ \epsilonup_{AB}\,{\bar{\Phi}}_{A'B'} \right)
\end{align*}
where $\{\bar{\Phi}_{A'B'}, \bar{\xi}_{A'}\}$ denote the complex conjugates of $\left\{\Phi_{AB}, \xi_{A}\right\}$ respectively and following the notation in \cite{penrose1987spinors}, $\epsilonup_{AB}=\epsilonup_{A'B'}$ is the $2\times 2$ symplectic matrix, $g^{\mu AA'}=(\mathbb{I},-\sigma_x,\sigma_y,-\sigma_z)/\sqrt{2}$ are the Infeld-van der Waerden symbols \cite{infeld1933}, $\sigma_i$ being the Pauli matrices.

A number of expressions simplify in this language:  Maxwell's equations become: $g^{\mu AA'}\partial_{\mu}{\Phi}_{AB}=0$ and the null condition (\ref{null_cond}): $\Phi_{AB}\Phi^{AB}=0$. The null field $\Phi_{AB}$ corresponding to the null GSF congruence $\xi_A$ is $\Phi_{AB}=\kappa\,\xi_{A}\,\xi_{B}$ where $\kappa$ is a complex scalar, chosen such that $\Phi_{AB}$ satisfies Maxwell's equations, and,  finally,  the condition for $\xi^\mu$ to be geodetic and shear-free (GSF)  simplifies to: $\xi^{A}\xi_{B}g^{\mu BB'}\partial_{\mu}\xi_{A}=0$. 

The Bateman field (\ref{emdef_bateman_basic}) corresponds to:
\begin{equation}
\xi_{A} = 
\begin{pmatrix}
\,\partial_{w}\,\bar{\alpha}\,\partial_{\bar{w}}\,\bar{\beta} - \partial_{\bar{w}}\,\bar{\alpha}\,\partial_{w}\,\bar{\beta}\,
\\
\,\partial_{\bar{w}}\,\bar{\alpha}\,\partial_{z}\,\bar{\beta} - \partial_{z}\,\bar{\alpha}\,\partial_{\bar{w}}\,\bar{\beta}\,
\end{pmatrix} 
\,, \: \kappa = \frac{{\rm i}} {\partial_{\bar{w}}\,\bar{\alpha}\,\partial_{z}\,\bar{\beta} - \partial_{z}\,\bar{\alpha}\,\partial_{\bar{w}}\,\bar{\beta}}
\label{Bateman_null_gsf}
\end{equation}

if $(\partial_{\bar{w}}\,\bar{\alpha}\,\partial_{z}\,\bar{\beta} - \partial_{z}\,\bar{\alpha}\,\partial_{\bar{w}}\,\bar{\beta}) \neq 0$, otherwise:
\begin{equation}
\xi_{A} = 
\begin{pmatrix}
\,\partial_{w}\,\bar{\alpha}\,\partial_{z}\,\bar{\beta} - \partial_{z}\,\bar{\alpha}\,\partial_{w}\,\bar{\beta}\,
\\
\, 0 \,
\end{pmatrix} 
\,, \: \kappa = \frac{{\rm i}} {\partial_{w}\,\bar{\alpha}\,\partial_{z}\,\bar{\beta} - \partial_{z}\,\bar{\alpha}\,\partial_{w}\,\bar{\beta}}
\label{Bateman_null_gsf_special_case}
\end{equation}
where $w=x+\rm{i}y$, and $(\bar{\alpha},\bar{\beta},\bar{w})$ denote the complex conjugates of $(\alpha,\beta,w)$ respectively. It is now explicitly seen that the entire family of null fields in (\ref{hfg_bateman}) corresponds to the above null GSF congruence $\xi_A$, with the complex scalar $\kappa$ rescaled by $\bar{h}$, the complex conjugate of $h$.

We list here two simple examples of this construction \cite{Bialynicki-Birula2004}: a circularly polarized plane wave traveling in the $+z$-direction and the Robinson congruence underlying the Hopfion solution. They arise from the following choices of $\alpha$ and $\beta$. For the plane wave: 
$\alpha = z-t,\,  \beta =x+\mathrm{i}y,\,  f=\mathrm{e}^{\mathrm{i}\alpha},\,g=\beta $, giving $\mathbf{F}^{\rm pw} =(\hat{x}+\mathrm{i}\hat{y})\mathrm{e}^{\mathrm{i}(z-t)}$. For the Hopfion we have instead: 
$\alpha =-d/b\,,\,\beta  =-\mathrm{i}a/(2b)\,, f =1/\alpha^2\,, g=\beta$ giving: $\mathbf{F}^{\rm hp} =d^{-3}(b^2-a^2,-\mathrm{i}( a^2+b^2),2ab)$ where $a =x-\mathrm{i}y,\,b=t-\mathrm{i}-z,\,d=r^2-(t-\mathrm{i})^2$. 

Equivalently: $\xi_A^{\rm pw}=(0,-1)\: , \,\kappa^{\rm pw}=-\mathrm{e}^{-\mathrm{i}(z-t)}\;$ and
\begin{equation}
\xi_A^{\rm hp}=(-\bar{b} , \bar{a} )\; , \;  \kappa^{\rm hp}=\bar{d}^{-3}  \label{hpf1}
\end{equation}
where $(\xi_A,\kappa)$ differ from those computed using ($\ref{Bateman_null_gsf}$) (with $\kappa$ rescaled by $\bar{h}$) by factors that leave the product $\Phi_{AB}=\kappa\,\xi_{A}\,\xi_{B}$ unchanged; $(\bar{a},\bar{b},\bar{d})$ are complex conjugates of $(a,b,d)$ respectively.

We now present a family of light-beam-like propagating solutions to Maxwell's equations in free space, in which the electric and magnetic fields encode torus knots and links which are preserved in time.
We proceed initially using Bateman's framework. We then describe in detail the knotted structure of the field lines, and compute the entire set of conserved currents, the helicity and charges for electromagnetism in free space for this family of knotted null fields.  

\begin{figure}
\includegraphics[width = \columnwidth]{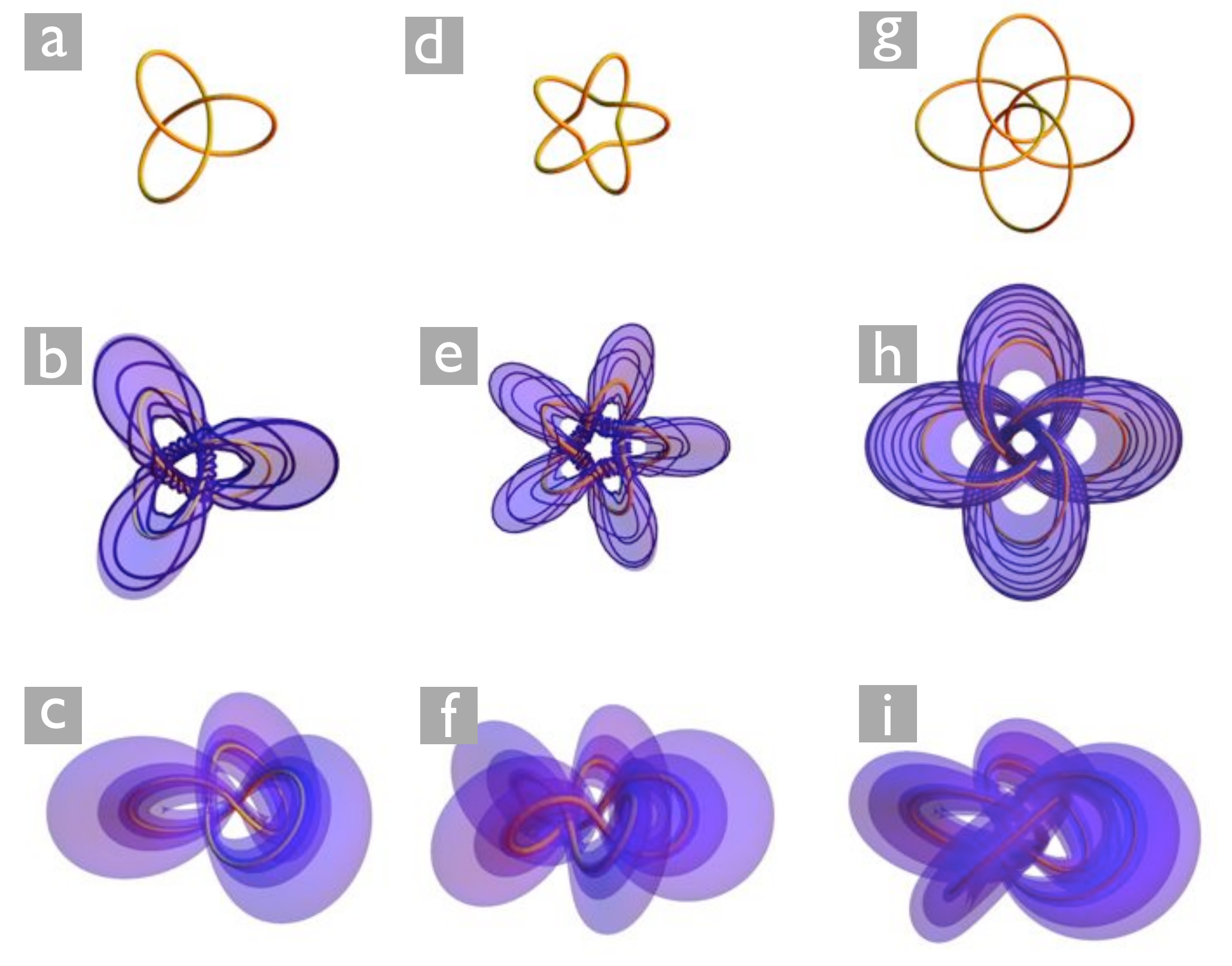}
\caption{{\bf Structure of magnetic field lines},
{\bf a-c}: Trefoil knots $(p=2,q=3)$, {\bf d-f}: Cinquefoil knots $(p=2,q=5)$, {\bf g-i}: $4$-Hopf linked rings $(p=2,q=2)$.  {\bf a},{\bf d},{\bf g}: Core (orange) field line(s) forming (a) a trefoil knot (d) a cinquefoil knot (g) $4$ linked rings. {\bf b},{\bf e},{\bf h}: Field line(s) (blue) wrapping around the core (orange) confined to a knotted torus (purple) enclosing the core. {\bf c},{\bf f},{\bf i}: Knotted nested tori (purple) enclosing the core, on which the field lines lie.}
\label{null_torus_knots1}
\end{figure}%

{\it Constructing knotted null electromagnetic fields --}
There is a natural connection between knots and singular points of complex maps from $\mathbb{S}^3$ to $\mathbb{C}$. This was used for example in recent work by Dennis et al. \cite{Dennis2010} to construct knotted optical vortices in light beams. In particular, it has been shown \cite{Brauner1928, milnor1969singular} that $u^{p}\pm v^{q}=0$ where $(u,v)\in \mathbb{C}^2$ is a pair of complex numbers such that $\vert u\vert^{2}+\vert v\vert^{2}=1$ (and hence they define coordinates on $\mathbb{S}^3$), gives a $(p,q)$ torus knot when $p$ and $q$ are coprime integers.

We note that the following choice of $(\alpha,\beta)$ in Bateman's construction:
\begin{equation}
\alpha = \frac{r^2 - t^2 - 1 + 2\mathrm{i}z }{r^2-(t-\mathrm{i})^2}, \quad\quad\beta = \frac{2(x-\mathrm{i}y)}{r^2-(t-\mathrm{i})^2}\,,\label{alphabeta_knotted}
\end{equation}%
which satisfies Eqs.~\eqref{alphabeta_def} and~\eqref{eq:nontrivial}, admits a natural interpretation as coordinates on $\mathbb{S}^3$ since $\vert\alpha\vert^{2}+\vert\beta\vert^{2}=1$ for any $t$. At $t=0$,  $(\alpha,\beta)=(u,v)$, the standard stereographic coordinates on $\mathbb{S}^3$ :
\begin{equation}
u=\frac{(r^2-1)+2\mathrm{i}z}{r^2+1}, \,\, v=\frac{2(x-\mathrm{i}y)}{r^2+1} \label{uv_def}
\end{equation}%

\begin{figure*}[!htb]
\includegraphics[width = 1.7 \columnwidth]{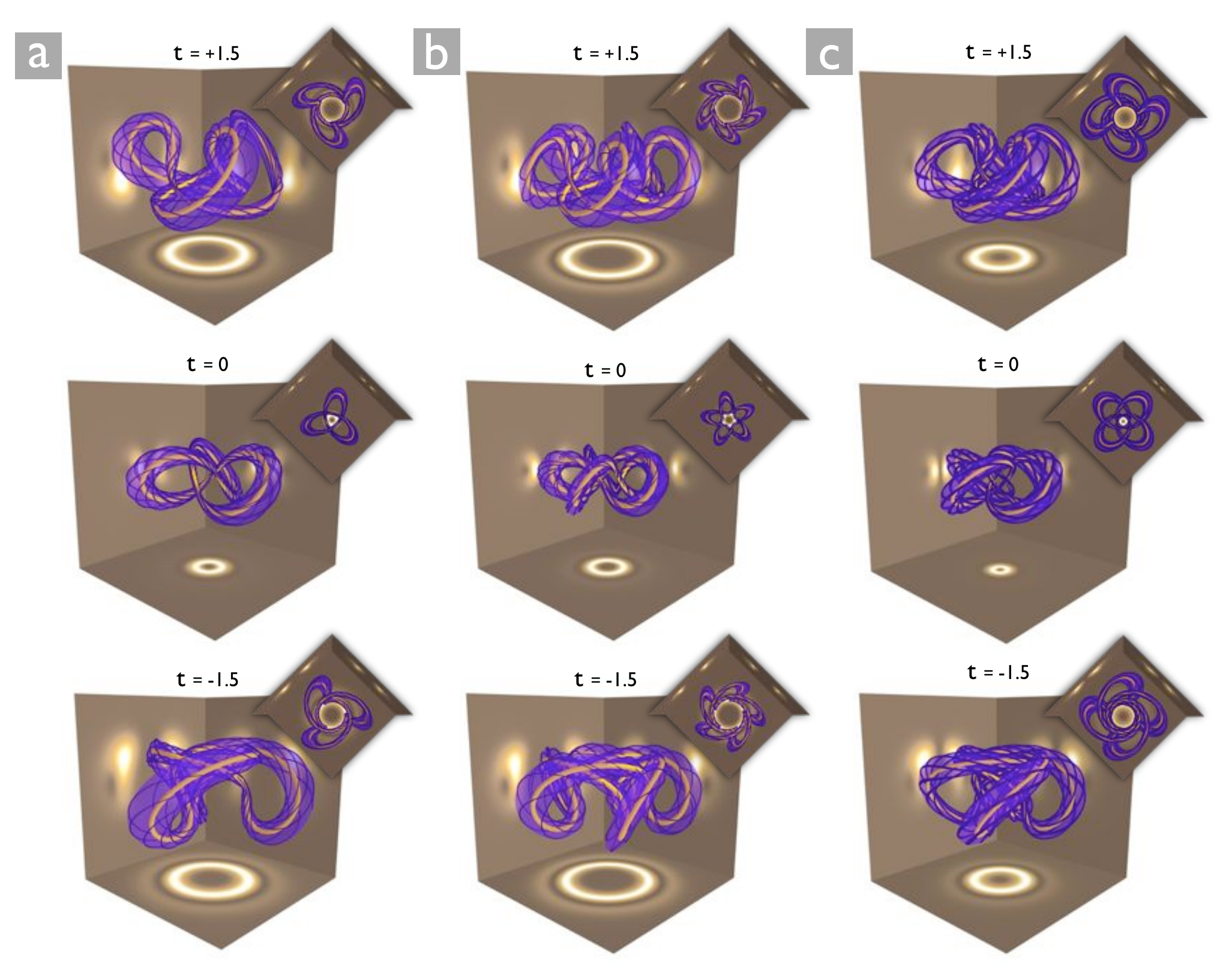}
\caption{{\bf Time evolution of magnetic field lines and energy density} for {\bf a}: the trefoil knot $(p=2,q=3)$,  {\bf b}: the cinquefoil knot $(p=2,q=5)$ and {\bf c}: the $4$ Hopf-linked rings $(p=2,q=2)$. Shown is the topology preserving, fluid-like evolution of the core field line(s) (orange), and field line(s) (blue) lying on tori (purple) enclosing the core.}
\label{null_torus_knots2}
\end{figure*}

Thus $\alpha^{p}\pm\beta^{q}=0$ encodes a singular line tied into a $(p,q)$ torus knot when $p,q$ are coprime integers. The simplest choice $f(\alpha,\beta)=\alpha^{p}$ and $g(\alpha,\beta)=\beta^{q}$ in (\ref{hfg_bateman}) gives rise to fields :
\begin{equation}
\mathbf{F}=\nabla\alpha^p\times\nabla\beta^q \label{knotted_bateman}
\end{equation}
whose electric and magnetic field lines are grouped into knotted and linked {\it tori}, nested one inside the other with $(p,q)$-torus knots at the core of the foliation. These being null electromagnetic fields, the topology of these knotted structures is preserved in time (See Fig.~\ref{null_torus_knots2}).

In the spinor formalism the fields arise from the same spinor as in (\ref{hpf1}) but different scaling factors $\kappa$:
\begin{equation}
\xi_{A}= (-\bar{b},\bar{a}), \,\, \kappa=4 p q\, \bar{\alpha}^{p-1}\; \bar{\beta}^{q-1}\bar{d}^{-3} \label{spinor_knotted_xi_kappa}
\end{equation}
where $\xi_A,\kappa$ have again been simplified leaving the product $\Phi_{AB}=\kappa\xi_A\xi_B$ unchanged. Thus, the entire family of knotted solutions is constructed by simply changing the scaling factor $\kappa$ in the Robinson congruence.

{\it Knotted structure of the field lines --}
We now examine the geometry of these solutions, starting with the magnetic field lines, illustrated in Fig.~\ref{null_torus_knots1}. The magnetic field lines organize around a set of {\it core} field lines which are closed and form $(p,q)$-torus knots or links. All magnetic field lines are tangent to nested tori enclosing these core field lines.

Specifically, the nested tori that the magnetic field lines lie on, are isosurfaces of  $\Psi_B=\mathrm{Re}\{\alpha^p \beta^q \}$, so that: $\mathbf{B}\cdot\nabla\Psi_B=0$. As $\Psi_{B}$ is varied from its maximum and minimum values $\left(\pm \sqrt{p^{p} q^{q} /(p+q)^{{p+q}}} \right)$ to zero, the corresponding isosurfaces are successively bigger tori, each enveloping the previous ones (see Fig.~\ref{null_torus_knots1}), until for $\Psi_{B}=0$ the isosurface extends to infinity. The {\it core} magnetic field lines occupy the loci $K_{B}^{\pm}$ of maxima and minima of $\Psi_B$:
\begin{align*}
K_{B}^{\pm} : (\alpha,\beta) = \frac{1}{\sqrt{p+q}}\Big(\mathrm{e}^{\mathrm{i}\left(q\theta+2\pi k/g \right)}\sqrt{p},\mathrm{e}^{\mathrm{i}\left(-p\theta -\pi/2q+2\pi k/g \pm\pi/2q \right)}\sqrt{q}\Big) 
\end{align*}
where $g=\gcd(p,q)$, $k\in\{0,1,\ldots,g-1\}$ and $\theta \in [0,2\pi)$ parametrizes the curve(s). The curves $K_{B}^{{\pm}}$ lie on a torus, winding $p$ times in the toroidal direction and  $q$ times in the poloidal direction, corresponding to $\beta$ and $\alpha$ changing phase by $-2\pi p$ and $2\pi q$ respectively, thus forming a $(p,q)$-torus knot for co-prime $p,q$.

We now describe the geometry of the core field lines $K_{B}^{{\pm}}$ for all possible values of the positive integers $(p,q)$.
\begin{enumerate}[i.]
\item When $(p\neq 1,q\neq 1)$ are coprime, the core field lines $K_{B}^{{\pm}}$ are a pair of linked $(p,q)$-torus knots. One of the core field lines $(K_{B}^{{+}})$, forming trefoil and cinquefoil knots is shown in Fig.~\ref{null_torus_knots1} (a,d).
\item When $p=1$ (or $q=1$), $K_{B}^{{\pm}}$ is a pair of linked rings with linking number $2q$ $(2p)$, sweeping around the torus $q$ ($p$)  times in the poloidal (toroidal) direction and once in the other direction.
\item In all other cases $g=\gcd(p,q)\neq 1$ can be factored out giving: $(p,q)=g\ast(\tilde{p},\tilde{q})$, where $(\tilde{p},\tilde{q})$ are coprime integers belonging to one of the two cases above. Then $K_{B}^{{\pm}}$ comprises of $2g$ linked $(\tilde{p},\tilde{q})$-torus knots if $(\tilde{p}\neq 1,\tilde{q}\neq 1)$ or $2g$ linked rings if either $\tilde{p}$ or $\tilde{q}$ is $1$ as shown for e.g. in Fig.~\ref{null_torus_knots1} (g).
\end{enumerate}%
All magnetic field lines are confined on the surface of tori (isosurfaces of $\Psi_B$) nested one inside the other about a common knotted core $K_{B}^{{\pm}}$, wrapping around the core, filling all space (see Fig.~\ref{null_torus_knots1}). Since the magnetic field is divergence-free and does not vanish on any isosurface $\Psi_B\neq 0$, it follows~\cite{Stern57} that all magnetic lines are either periodic or quasi-periodic on each toroidal surface. 

The electric field lines have exactly the same structure, rotated in space about the $z$-axis by $\pi/(2q)$. The corresponding knotted tori that the electric field lines lie on, are isosurfaces of $\Psi_E=\mathrm{Im}\{\alpha^p \beta^q \}$ and the core field lines are given by $K_{E}^{{\pm}}:\Psi_E = \pm\sqrt{p^{p} q^{q} /(p+q)^{{p+q}}}$.

This description of the knotted structure of the electric and magnetic field lines is in terms of the time-dependent $(\alpha,\beta)$ and holds true for all time, conforming to our expectation of the field line structure of a null field being preserved with time (as seen in Fig.~\ref{null_torus_knots2}). Surprisingly, the case $p=q=1$ gives the Hopfion solution described earlier in Fig.~\ref{Hopfion_fig}. The field lines conform to our earlier description and lie on isosurfaces of $\Psi_B$ $(\Psi_E)$, but all field lines are closed and are linked with every other field line.

To further characterize the physical properties of this family of knotted null fields, we compute the helicity and the full set of conserved quantities \cite{Irvine2008} corresponding to the known (conformal) symmetries of electromagnetism in free space. The non-vanishing currents and charges normalized by the energy are:\\
Magnetic helicity $\mathcal{H}_m\;=\;$ Electric helicity $\mathcal{H}_e \;=\;\frac{1}{p+q}\,,$\\
Momentum $\mathbf{P}\;=\;$ SCT current $\mathbf{v}\;=\;\left(0,0,\frac{-p}{p+q}\right)\,,$ \\ 
Angular momentum $\mathbf{L}\;=\;\left(0,0,\frac{q}{p+q}\right)\,$.

{\it Summary --} The solutions presented here extend the space of known solutions beyond the  Hopfion, by encoding an entire family of both knots and links that are preserved under time evolution. Many open questions remain on the space of knotted states, such as whether solutions with each and every field line knotted and preserved by the time evolution exist with topology different from the Hopf fibration (e.g. a Seifert foliation of $\mathbb S^3$). Beyond electromagnetism it remains an open question whether  similar explicit solutions be found for nonlinear evolutions such as the Euler flow of ideal fluids. From a dynamical systems perspective, it may be interesting to explore the role of the  invariant tori in the solutions we present and the conditions for which Bateman's construction give rise to electric and magnetic fields with a first integral.  Finally, if realized in experiment, can these structures be imprinted on matter such as plasmas or quantum fluids?  

{\bf Acknowledgements: } {\small The authors acknowledge interesting discussions with Mark Dennis, Ricardo Mosna and Radmila Sazdanovic and the hospitality of the KITP during the program ``Knotted Fields''. WTMI and DPS further acknowledge the hospitality of the Newton Institute during the program ``Topological fluid dynamics''. D.P.-S. is supported by the Spanish MINECO under grant no. MTM2010-21186-C02-01, the ICMAT Severo Ochoa project SEV-2011-0087 and the Ram\'on y Cajal program. W.T.M.I.  acknowledges support from the   A.P. Sloan Foundation through a Sloan fellowship, and  the  Packard Foundation through a Packard fellowship. }
\bibliography{emknots}

\end{document}